\documentclass[11pt,a4paper]{article}

\usepackage{amsmath,amssymb,bm}
\usepackage{graphicx}
\usepackage{hyperref}
\usepackage{tikz}
\usepackage{authblk}

\newcommand{\bc}{\mathbf{c}}
\newcommand{\bq}{\mathbf{q}}
\newcommand{\grad}{\boldsymbol{\nabla}}

\newcommand{\bJM}{\mathbf{J}_{M}}
\newcommand{\bJE}{\mathbf{J}_{E}}
\newcommand{\Pran}{\mathrm{Pr}}

\newcommand{\avg}[1]{\langle #1 \rangle_0}

\newcommand{\XiTwo}{\Xi^{(2)}}

\begin{document}

\title{Conductive Heat Flux Driven by a Pressure Gradient in Non-Maxwellian Reference States}

\author[1,2,3]{Jae Wan Shim}

\affil[1]{Extreme Materials Research Center, Korea Institute of Science and Technology, 5 Hwarang-ro 14-gil, Seongbuk, Seoul, 02792,  Republic of Korea}

\affil[2]{Climate and Environmental Research Institute, Korea Institute of Science and Technology, 5 Hwarang-ro 14-gil, Seongbuk, Seoul, 02792,  Republic of Korea}

\affil[3]{Division of AI-Robotics, KIST Campus, University of Science and Technology, 5 Hwarang-ro 14-gil, Seongbuk, Seoul, 02792,  Republic of Korea}

\emergencystretch=2em

\maketitle
\begin{abstract}
Standard Navier--Stokes--Fourier theory and Maxwellian-based Grad 13-moment closures yield no independent pressure-gradient driving of the conductive heat flux in an isothermal, single-component gas in the hydrodynamic (small-Knudsen) regime. This absence is specific to the Maxwellian local-equilibrium weight. We show that when the closure is constructed about a generalized class of isotropic non-Maxwellian reference weights with finite fourth moment---characterized by a single shape parameter (a kurtosis-like moment ratio) that deforms the distribution continuously away from a Maxwellian---the small-Knudsen constitutive reduction retains a bulk pressure-gradient (barothermal) contribution to the conductive heat flux. This mechanism predicts pressure-driven conduction as a direct kinetic signature of non-Maxwellian equilibrium moment structure.
\end{abstract}

\section{Introduction}
Standard Navier--Stokes--Fourier (NSF) theory contains no independent bulk pressure-gradient ($\nabla p$) driving in the constitutive law for the \emph{conductive} heat flux $\bq$ of a single-component gas in the hydrodynamic (small-Knudsen) regime: at leading order, $\bq$ responds to temperature gradients alone~\cite{ChapmanCowling1970,deGrootMazur1962}.
The same structure emerges from Maxwellian-based Grad's 13-moment method: the leading-order constitutive reduction admits no independent bulk pressure-gradient driving of $\bq$~\cite{Grad1949CPAM,Struchtrup2005}.

We show that replacing the Maxwellian local-equilibrium weight, the maximizer of Boltzmann--Gibbs entropy,
by a generalized equilibrium weight obtained by maximizing the Havrda--Charv\'at entropy~\cite{Havrda1967}
(later reintroduced by Tsallis~\cite{Tsallis1988}) yields, in the Grad-13 small-Knudsen constitutive reduction, an additional \emph{bulk} barothermal contribution to the conductive heat flux, $\bq\propto\nabla p$, already at leading order.

Our starting point is a class of isotropic local reference weights that deform continuously away from a Maxwellian through a single shape parameter. Such generalized non-Maxwellian distributions arise in many circumstances~\cite{DouglasBergaminiRenzoni2006PRL,PierrardLazar2010SSR,Leubner2000PSS,LatoraRapisardaTsallis2001PRE}, including the one-particle marginal associated with a fixed-energy constraint~\cite{ScalasEtAl2015PRE} and a local stationary state of generalized (non-Boltzmann) kinetic dynamics (e.g.\ nonlinear Fokker--Planck models). In the microcanonical case, the same marginal can be obtained from Havrda--Charv\'at (Tsallis) entropy maximization, which also yields an explicit relation between the corresponding entropic index and the system's particle number~\cite{shim2018arxiv, shim2020entropy}.
When Grad's 13-moment closure is constructed with these generalized non-Maxwellian equilibrium weights, the small-Knudsen constitutive reduction yields an additional \emph{bulk barothermal} contribution to the conductive heat-flux law, $\bq\propto\nabla p$, which can be isolated under isothermal conditions ($\nabla\theta=0$). In this formulation the coefficient of the barothermal term is controlled by a single kurtosis-like equilibrium moment ratio, so a concrete microcanonical realization can be viewed as a parameter-free instance of a more general mechanism.

Note that this bulk constitutive effect should be distinguished from boundary-mediated rarefaction phenomena such as thermal transpiration and Knudsen effusion, which rely on wall-induced nonequilibrium and are not the origin of the present $\nabla p$ coupling~\cite{Sone2007,Bird1994}.
For pressure-driven configurations it is also useful to recall that the measurable energy flux can be decomposed as
\begin{equation}
  \bJE=\bq+h\,\bJM,
  \label{eq:JE_decomp_intro}
\end{equation}
where $\bJE$ is the total energy flux, $\bJM$ is the mass flux, and $h$ is the specific enthalpy~\cite{deGrootMazur1962,ChapmanCowling1970}.
This decomposition clarifies how a pressure-gradient contribution in $\bq$ may be identified against the advective enthalpy transport $h\,\bJM$ in pressure-driven realizations.

\section{Generalized Grad-13 closure and the barothermal effect}
Following Grad's 13-moment method~\cite{Grad1949CPAM,Struchtrup2005}, we formulate a Grad-type 13-moment closure and generalize it by constructing the polynomial basis orthogonal with respect to an arbitrary isotropic local reference weight $f^{(0)}$.
Isotropy preserves the canonical tensorial structure of the Grad equations, while the scalar transport coefficients are renormalized by the moment structure of $f^{(0)}$.
For an isotropic weight $f^{(0)}(\bc)$ (with peculiar velocity $\bc$), the equilibrium input relevant to the barothermal coupling can be reduced to a single kurtosis-like ratio,
\begin{equation}
\XiTwo \equiv \frac{\avg{c^4}}{\theta\,\avg{c^2}},
\label{eq:Xi_def}
\end{equation}
where $\avg{\cdot}$ denotes averaging with respect to $f^{(0)}$.
We write $\theta \equiv k_B T/m$, where $k_B$ is Boltzmann's constant, $T$ the temperature, and $m$ the particle mass.

Linearizing about a quiescent state and performing the small-Knudsen constitutive reduction, we obtain the generalized leading-order heat-flux law
\begin{equation}
\bq=-\kappa\left(A\,\grad\theta+B\,\frac{1}{\rho}\grad p\right)
+O(\mathrm{Kn}^2),
\label{eq:generalized_Fourier}
\end{equation}
where $\rho$ is the mass density and $\mathrm{Kn}$ the Knudsen number.
We use the Maxwellian reference prefactor $\kappa=\frac{5}{2}\tau_q p$ (with $\tau_q$ the heat-flux relaxation time)~\cite{Grad1949CPAM,Struchtrup2005}, so that the effective Fourier coefficient is $\kappa A$.

The coefficients $A$ and $B$ are determined solely by the deviation of $\XiTwo$ from its Maxwellian value:
\begin{equation}
A=\frac{\XiTwo}{5},\quad
B=\frac{\XiTwo-5}{5}.
\label{eq:A_coeff}
\end{equation}

For a Maxwellian weight in three dimensions, the Gaussian identity $\avg{c^4}=5\,\theta\,\avg{c^2}$ implies $\XiTwo=5$ and hence $B=0$ in Eq.~\eqref{eq:A_coeff}.
Thus, Maxwellian-based Grad-13 yields $A=1$ and $B=0$, whereas any isotropic reference weight with finite $\avg{c^4}$ and $\XiTwo\neq 5$ yields a nonzero bulk $\nabla p$ contribution to $\bq$ at the same hydrodynamic order.

\section{Realizations: from compact support to heavy tails}
We illustrate the mechanism using the isotropic $q_s$-Gaussian (Tsallis) family of local reference weights, which maximizes the Havrda--Charv\'at entropy under fixed first and second velocity moments~\cite{Havrda1967,Tsallis1988}:
\begin{equation}
f^{(0)}_{q_s}(\bc)\propto
\biggl[1-\frac{1-q_s}{2\beta^{2}}\,c^{2}\biggr]_+^{\frac{1}{1-q_s}},
\quad
\beta>0,
\label{eq:q_gaussian_def_realizations}
\end{equation}
where $[y]_+ := \max\{y,0\}$ and $c^2=\bc\cdot\bc$. We consider the isotropic (zero-mean) case $\avg{\bc}=\mathbf 0$.
The scale $\beta$ is fixed by the kinetic-temperature (second-moment) constraint. In three dimensions we impose $\avg{c^2}=3\theta$. This fixes $\beta=\beta(q_s,\theta)$. For $q_s<1$ the support is compact, whereas for $q_s>1$ the distribution is heavy-tailed.
The Maxwellian limit is recovered as $q_s\to 1$.

In three dimensions, a direct evaluation yields the kurtosis-like ratio
\begin{equation}
\XiTwo(q_s) = 5\,\frac{7-5q_s}{9-7q_s},
\label{eq:Xi_qs}
\end{equation}
valid for the compact-support side $q_s<1$ and for the heavy-tailed side $1<q_s<9/7$ where the fourth moment is finite.
Accordingly, $q_s<1$ gives $\XiTwo<5$ and $q_s>1$ gives $\XiTwo>5$, implying a sign change of the barothermal coefficient $B=(\XiTwo-5)/5$ across $q_s=1$.

\subsection{Microcanonical realization: the fixed-energy-shell one-particle marginal}
As a clean realization, consider an isolated ideal gas on a fixed total kinetic-energy shell.
Its exact one-particle marginal corresponds to the compact-support case ($q_s < 1$).
In three dimensions, the entropic index $q_s$ is related to particle number $N$ by $q_s=(3N-7)/(3N-5)$~\cite{shim2018arxiv, shim2020entropy}.
Inserting this relation into Eq.~\eqref{eq:Xi_qs} reproduces the microcanonical result:
\begin{equation}
\XiTwo_N=\frac{15N}{3N+2}.
\label{eq:Xi_result}
\end{equation}
In particular, $\XiTwo\to 5$ as $q_s\to 1$ (equivalently $N\to\infty$).
Substituting $\XiTwo=\XiTwo_N$ into Eq.~\eqref{eq:A_coeff} yields
\begin{equation}
B_N=-\frac{2}{3N+2} \approx -\frac{2}{3N}.
\label{eq:AN_result}
\end{equation}
The sign $B_N<0$ implies that, under isothermal pressure gradients, the conductive contribution aligns with $+\nabla p$.

\subsection{Heavy-tailed realization: a $q_s>1$ stationary state}
To emphasize that the barothermal mechanism is not tied to compact support, we also note a complementary heavy-tailed realization with $q_s>1$.
Such power-law stationary states arise in generalized Fokker--Planck dynamics~\cite{Lutz2003PRA,Borland1998PLA,Borland1998PRE}.
As a representative velocity-space model, consider
\begin{equation}
\partial_t f
= \gamma \nabla_{\bc}\!\cdot(\bc f)
+ D \nabla_{\bc}^2\!\left(f^{\,2-q_s}\right).
\label{eq:fp_q}
\end{equation}
Here $f(\bc,t)$ is the one-particle velocity distribution, $\gamma>0$ is a linear friction (drift) rate, and $D>0$ is a generalized diffusion coefficient in velocity space.
At stationarity ($\partial_t f=0$), the drift term balances the nonlinear diffusion, yielding an isotropic $q_s$-Gaussian of the form~\eqref{eq:q_gaussian_def_realizations} for an appropriate second-moment (temperature) constraint.

In three dimensions, finiteness of the fourth moment requires $q_s<9/7$ (and in particular $1<q_s<9/7$ on the heavy-tailed side); as $q_s\to(9/7)^-$ the fourth moment diverges and the present Grad-13 input ceases to exist.
Within this domain, Eq.~\eqref{eq:Xi_qs} applies, and $\XiTwo\to 5$ as $q_s\to 1$.
Using Eq.~\eqref{eq:A_coeff}, the barothermal coefficient is
\begin{equation}
B(q_s)=\frac{2(q_s-1)}{9-7q_s}.
\label{eq:B_heavy}
\end{equation}
For $1<q_s<9/7$, both numerator and denominator are positive, so $B>0$.
Thus, under isothermal pressure gradients, the conductive contribution aligns with $-\nabla p$.

\section{Observability and competition with advective enthalpy transport}
To leading order in the small-Knudsen expansion, Eq.~\eqref{eq:generalized_Fourier} predicts an isothermal pressure-driven \emph{conductive} heat flux:
\begin{equation}
\bq=-\kappa\,B\,\frac{1}{\rho}\,\grad p
\qquad(\grad\theta=0).
\label{eq:isothermal_baro_q}
\end{equation}
In pressure-driven flows, however, the measured energy flux $\bJE$ in Eq.~\eqref{eq:JE_decomp_intro} also contains the advective enthalpy term $h\,\bJM$~\cite{deGrootMazur1962}.
To quantify when the conductive barothermal contribution can be resolved against this background, define
\begin{equation}
\mathcal{C}\equiv\frac{|\bq|}{|h\,\bJM|}.
\label{eq:C_def}
\end{equation}
Using a minimal Poiseuille/Darcy parameterization $\bJM\sim-(\rho L_h^2/\mu)\,G\,\grad p$ with hydraulic length $L_h$, viscosity $\mu$, and geometry factor $G=O(1)$, one obtains the scaling estimate
\begin{equation}
\mathcal{C}\sim \frac{|B/A|}{ G\,\Pran}\left(\frac{\nu}{L_h\sqrt{\theta}}\right)^2,
\qquad
\nu\equiv\mu/\rho,
\label{eq:C_scaling}
\end{equation}
where $\Pran$ is the Prandtl number, defined using the effective Fourier coefficient $\kappa A$.
Although the constitutive reduction assumes $\mathrm{Kn}\ll 1$, the ratio \eqref{eq:C_scaling} increases with rarefaction and can become appreciable already at moderate rarefaction (slip/early transitional regimes), where higher-order corrections remain controlled~\cite{Sone2007}.

Thus the barothermal conductive contribution is most accessible in mesoscopic channels (small $L_h$) and/or when the equilibrium weight is sufficiently non-Gaussian in the sense that $|B/A|$ is not too small.

\section{Conclusion}
We have identified a bulk constitutive mechanism by which a pressure gradient drives a \emph{conductive} heat flux in an isothermal, single-component gas within a Grad-13 framework. 
In contrast to NSF theory and the standard Maxwellian-based Grad-13 closure, where the leading-order constitutive reduction yields no independent bulk $\nabla p$ driving of $\bq$, the present result shows that this cancellation is not dictated by general invariance requirements, but is tied to the Maxwellian moment identity $\XiTwo=5$ in three dimensions~\cite{ChapmanCowling1970,Grad1949CPAM,Struchtrup2005,Curie1894,deGrootMazur1962}. 
Replacing the Maxwellian local-equilibrium weight (the Boltzmann--Gibbs entropy maximizer) by generalized local equilibria obtained, for instance, from Havrda--Charv\'at entropy maximization~\cite{Havrda1967,Tsallis1988} produces, under the same small-Knudsen constitutive reduction, the generalized Fourier law~\eqref{eq:generalized_Fourier} with a barothermal coefficient $B=(\XiTwo-5)/5$. 
Hence, whenever $\XiTwo\neq 5$, pressure-driven conduction persists already at leading order, even for $\grad\theta=0$.

A concrete realization is provided by the microcanonical fixed-energy-shell one-particle marginal, for which $\XiTwo=\XiTwo_N$ and therefore $B=O(1/N)$, with $B<0$ implying $\bq$ aligned with $+\grad p$ in isothermal pressure-gradient configurations. 
Equivalently, the same marginal can be parameterized by the nonadditive entropic index $q_s$, via $\XiTwo=\XiTwo(q_s)$, linking the barothermal transport coefficient directly to an underlying entropy-based characterization of local equilibrium~\cite{shim2018arxiv,shim2020entropy}. 
Finally, since experiments access the total energy flux $\bJE=\bq+h\,\bJM$, the observability of the barothermal \emph{conductive} signal against advective enthalpy transport is governed by $\mathcal C$ in Eq.~\eqref{eq:C_def}, with the scaling estimate~\eqref{eq:C_scaling} indicating enhanced accessibility in mesoscopic channels (small $L_h$) and/or for sufficiently non-Gaussian equilibria (i.e., $|\XiTwo-5|$ not too small), provided the relevant low-order moments are finite (e.g.\ for the $q_s$-Gaussian family in 3D, $q_s<9/7$). 
The predicted pressure-driven conductive heat flux thus provides a direct kinetic signature of non-Maxwellian equilibrium moment structure.

\section*{Acknowledgments}
This work was partially supported by the KIST Institutional Program.

\bibliographystyle{unsrt}
\bibliography{references}

\end{document}